\input{aipcheck}

\documentclass[
    ,final            % use final for the camera ready runs
  ]
  {aipproc}

\layoutstyle{6x9}

\newcommand{\gsim}{\raisebox{-0.6ex}{$\stackrel{{\displaystyle>}}{\sim}$}}
\newcommand{\pdot}{$\dot{\rm P}$}
\newcommand{\msun}{M$_{\odot}$}
\newcommand{\rsun}{R$_{\odot}$}

\begin{document}

%\title{Searching for Evolved Planetary Systems with the Timing Method:
%Past, Present and Future}

\title{The Potential of the Timing Method to Detect Evolved Planetary Systems}



\author{Roberto Silvotti}{address={INAF-Osservatorio Astronomico di Torino, 
strada dell'Osservatorio 20,\\ 10025 Pino Torinese, Italy}}

\author{R{\'o}bert Szab{\'o}}{address={Konkoly Observatory of the Hungarian Academy 
of Sciences, 
Konkoly Thege Mikl\'os \'ut 15-17, 
1121~Budapest, Hungary}}

\author{Pieter Degroote}{address={Instituut voor Sterrenkunde, K.U. Leuven,
Celestijnenlaan 200D, 3001 Leuven, Belgium}}

\author{Roy H.\ \O stensen}{address={Instituut voor Sterrenkunde, K.U. Leuven,
Celestijnenlaan 200D, 3001 Leuven, Belgium}}

\author{Sonja Schuh}{address={Georg-August-Universit\"at G\"ottingen, Institut 
f\"ur Astrophysik, 
Friedrich-Hund-Platz 1, 
37077~G\"ottingen, Germany}}

\begin{abstract}
The timing method, using either stellar pulsations or eclipse timing of close binaries
as a clock, is proving to be an efficient way to detect planets around stars
that have evolved beyond the red giant branch.
In this article we present a short review of the recent discoveries
and we investigate the potential of the timing method using data both from 
ground-based facilities as well as from the Kepler and CoRoT space missions.
\end{abstract}

\maketitle

%%%%%%%%%%%%%%%%%%%%%%%%%%%%%%%%%%%%%%%%%%%%
%% MAINMATTER
%%%%%%%%%%%%%%%%%%%%%%%%%%%%%%%%%%%%%%%%%%%%

\section{Introduction}

The large majority of the almost 500 extrasolar planets detected so 
far\footnote{494 from the Extrasolar Planets Encyclopaedia (%
\url{http://exoplanet.eu/}) on October 12th, 2010.}
orbit main sequence (MS) stars, and these planets were detected mostly using 
radial velocities (RVs) or transits.
For various reasons, not only technical, current exoplanet research is strongly
oriented towards MS stars, with the main goal of finding rocky analogs of our
Earth, and towards understanding planetary system formation, while the late-stage 
evolution after the MS remains poorly understood.
The first attempts to model the evolution of the planetary systems after
the red giant phase 
are very recent (Villaver \& Livio 2009 and ref.\ therein), and in particular lack 
observational constraints.
The two main methods to detect exoplanets, RVs and transits, are not efficient 
for compact stars with high gravities and small radii, such as the 
extreme horizontal branch stars and the white dwarfs.
An alternative method for these stars is given by the timing method,
in analogy with the first pulsars' planet discoveries \cite{Wolszczan+1992}.
This method is based on the reflex motion of the parent star due to the 
companion, which changes periodically the star-observer distance, causing a 
delay (or advance) on the arrival time of the photons.
In order to use the timing method, what we need is a precise astrophysical
clock.
For the pulsars' planets, this is given by the pulsar's radio signal
modulated by the fast and stable rotation of the neutron star.
But other potentially precise clocks exist in astrophysics and are given, 
for example, by stellar pulsation timing or the eclipse timing in close
binary systems.
An early example of application of the timing method is given by
Papar\'o et al.\ (1988), who calculated the orbital elements of an unseen 
stellar companion to the Delta Scuti star SZ Lyncis using the O--C technique
(Sterken 2005) on the pulsation timing.
The method has also been discussed for pulsating white dwarfs by 
Kepler et al.\ (1991) and Provencal (1997).

The next sections contain an introduction to the field of post-RGB
planetary systems and a short review of the recent discoveries,
respectively.
Then we investigate the potential of the timing method to find 
new planetary systems around evolved stars using pulsation or eclipse 
timing, with particular attention to the ongoing space missions CoRoT and 
Kepler. Finally, in the last section, a few conclusions are given and 
future perspectives are discussed.

\section{Post-RGB evolution of planetary systems}

During the red giant (RG) expansion of a star, two main opposite phenomena 
influence the orbits of potential planetary companions: mass loss from the star,
which lowers the binding energy and hence tends to cause an outward drift of
the planet, and tidal effects, which tend to attract the planet close 
to the star and eventually inside the envelope.
If the planet enters the stellar envelope, it is not clear in which conditions it 
can survive and what can be the influence on the subsequent evolution of the 
star.
On the other hand, we know that during the RG phase a small fraction of stars,
of the order of 1\%, loose almost all their envelope and are left with 
a very thin H envelope during their core He burning phase.
These are the so-called subdwarf B (or sdB) stars, that are located on the 
extreme horizontal branch (EHB), at high effective temperatures (\gsim 
20,000\,K).
At least 50\% of them are in binary systems and in this case the formation of
an sdB star can be explained in terms of binary evolution 
(Han et al.\ 2002, 2003).
However, for the remaining fraction of apparently single sdB stars, it is more
difficult to explain why the huge mass loss during the RGB occurred
(see review by Heber 2009).
In this context planets might play a role \cite{Soker1998}.

Similar phenomena of planetary migration, planet engulfment and planet
accretion from stellar ejected material can happen also during the second large
expansion of a star, when it enters the asymptotic giant branch and the thermal 
pulses phase, and during the planetary nebula (PN) ejection.

In order to achieve consistent modelling of all these complex phenomena, tests 
against observational constraints are essential.
To detect planets around white dwarfs is (or more exactly will be) a key step 
to study the late-stage evolution of planetary systems.
Equally important is to identify planetary systems in previous shorter 
and more rare evolutionary phases like the RGB or the horizontal branch.
Only in this way can we try to disentangle the various phenomena that 
changed the configuration of the planets and their orbits, with the final 
goal of being able to trace back the entire history of a planetary system 
from the final white dwarf configuration to the original protoplanetary disk.
What we need first is to find a sufficient number of planetary systems 
orbiting post-RGB stars and white dwarfs, covering a wide range of 
orbital parameters and planetary masses.

Concerning white dwarfs, what we know already is that some of
these stars are surrounded by metal-rich debris disks:
at least 17 white dwarfs show circumstellar dust at $\sim$ 1,000~K
(Reach et al.\ 2005 and ref.\ therein, Farihi et al.\ 2010) and about 
20\% of them show also Ca\,{\sc ii} emission from a hotter gaseous disk 
(G\"ansicke et al.\ 2006, 2010).
These disks are generally very close to the white dwarf, within about 1 \rsun.
%from the white dwarf.
Recently, also much more distant and cooler debris disks have been detected 
(Su et al.\ 2007, Chu et al.\ 2010 and Bil\'ikov\'a et al.\ 2010), 
which are probably the final evolution of the solar system's 
Kuiper belt analogues.
In other words, we don't know yet how many white dwarfs host planets and at 
which distances from the star, but we know that at least the material to build 
planets is there.

Another interesting possibility is that second (or even third) 
generation planets can form from the stellar material ejected during 
stellar evolution (Perets et al.\ 2010). But again, to confirm these
theoretical scenarios, we need observations and statistics.

\section{Recent detections of post-RGB planetary systems}

From the observational side, the late-stage evolution of the planetary systems
is a very recent field and only in the last three years a few first post-RGB 
planetary systems were detected.
They belong mainly to two classes:

\begin{enumerate}

\item
Extreme horizontal branch hot subdwarfs B:\\\
V391 Peg~b \cite{Silvotti+2007},\\
HW Vir~b and c \cite{Lee+2009},\\
HS\,0705+67003~b \cite{Qian+2009},\\
HD\,149382~b (Geier et al.\ 2009, but see also Jacobs et al.\ 2010).
% Jacobs et al.\ these proceedings!!!

\item
Cataclysmic variables:\\
NN Ser~b and c
(Qian et al.\ 2009b, Beuermann et al.\ 2010a, Hessman et al.\ 2010),\\
DP Leo~b (Qian et al.\ 2010a, Beuermann et al.\ 2010b) and\\ 
QS Vir~b (Qian et al.\ 2010b).

\end{enumerate}

We note that the issue of nomenclature for circumbinary planets, in 
contrast to the nomenclature for planets around components of visually 
resolved binaries, is unsolved. On recommendation by the \emph{A\&A} editor,
Beuermann et al.\ (2010a, 2010b) follow the rule by which any unseen companions,
whether 
stellar or sub-stellar, are designated by lowercase letters. They hence
have introduced (ab) for the two components of the central binary and start 
numbering the third bodies with the letter c, resulting in the planet 
designations NN Ser~(ab)~c, NN Ser~(ab)~d, and DP Leo~(ab)~c.

Another interesting new planet candidate was presented during
this meeting by Setiawan et al.\ 2010a
(a 1.3 M$_{\rm Jup}$ planet at 0.116 AU from a red HB star of likely 
extragalactic origin).

In Table~1 we summarize the list of known post-RGB planets, including the
pulsars' planets and a few brown dwarfs (BDs) near the planet mass limit.
We include also the brown dwarf WD\,0137$-$349\,b because its very small
orbital distance, only 0.375 \rsun, implies that the BD was engulfed 
by the red giant progenitor of the white dwarf (Maxted et al.\ 2006).
%\clearpage
As one can see, most of the detections reported in Table~1 were obtained 
through timing measurements.

\begin{table}[!t]
\begin{tabular}{l l@{}r@{}c@{}l @{~} r@{}c@{}l r@{}c@{}l @{\rule{5mm}{0mm}} l l l l}
\hline
%  & \tablehead{1}{r}{b}{Small\tablenote{2-9 retail outlets}\\multiple}
%  & \tablehead{1}{c}{b}{T$_{eff}$ (K)\tablenote{DA WDs with $\log${\it 
%  g}=8.0}\\}  
%  & \tablehead{1}{c}{b}{vol$_{50, 13}$~[pc$^{3}$]\tablenote{d$\le$50, 
%  V$\le$13}\\}
    \tablehead{1}{l}{b}{Planet}
  & \tablehead{4}{c}{b}{$\mathbf{M\,\sin\,{i}}$~[M$_ {\rm Jup}$]}
  & \tablehead{3}{c}{b}{$\mathbf{a}$ [AU]}
  & \tablehead{3}{c}{b}{$\mathbf{P}$~[yr]}
  & \tablehead{1}{l}{b}{$\mathbf{e}$}
  & \tablehead{1}{l}{b}{Evolut. phase}
  & \tablehead{1}{l}{b}{Detection method}
  & \tablehead{1}{l}{b}{References} \\
\hline
PSR\,1257$+$12\,b  & $M\simeq$&&&6\,e$-$5 & 0&.&19 & 25&.&262\hfill d  & 0      & pulsar 	& timing (radio signal)& Wolszczan \& Frail 1992 \\ 
PSR\,1257$+$12\,c  & $M\simeq$&0&.&014 & 0&.&36 & 66&.&5419~\hfill d & 0.0186 &         	& timing (radio signal)& Konacki \& Wolszczan 2003\\
PSR\,1257$+$12\,d  & $M\simeq$&0&.&012 & 0&.&46 & 98&.&2114~\hfill d & 0.0252 &        	& timing (radio signal)& \\
\hline
PSR\,B1620$-$26\,b && 2&.&5   &23&& 	       & 100&&	    &	     & pulsar+WD	& timing (radio signal)& Thorsett et al.\ 1993 \\
                   &&&&	      &&&	              && &	    &	     & in globular cluster&                    & Sigurdsson et al.\ 2003\\
%\hline
%GD\,165\,b       &&&&	 &&& $\sim$120   &	    &&&	     & WD 		& IR excess	       & Becklin \& Zuckerman 1988\\
%                 &&&&	 &&&	       &	    &&&	     &  		&		       & Zuckerman \& Becklin 1992\\
%\hline
%GD\,1400\,b      &&&&	 &&&	       &	    &&&	     &  		&		       & Farihi \& Christopher 2004\\
%                 &&&&	 &&&	       &	    &&&	     &  		&		       & Farihi et al.\ 2005\\
%\hline
%2MASS J0030$-$3739 &&&& 0.07-0.08 \msun &&&  &	    &&&	     &  		&		       & Day-Jones et al.\ 2008\\
\hline
WD\,0137$-$349\,b & $M\simeq$&55&& & 0&.&375 \rsun & 1&.&927\hfill h   &   & WD		& RVs		       & Maxted et al.\ 2006\\
\hline
%HD\,13445=Gl\,86 &&&& 4.01  &&& 0.11        & 15&.&77\,d   & 0.046  & MS+WD at 19 AU	&		       & Queloz et al.\ 2000\\
%\hline
%HD\,147513       &&&& 1	 &&& 1.26        & 1&.&48	    & 0.52   & pMS+WD		&		       & \\
%                 &&&&	 &&&	       &	    &&&	     &  		&		       & Holberg et al.\ 2002\\
%\hline
%HD\,27442        &&&& 1.28 &&& 1.18        & 1&.&160      & 0.07   & MS+WD at 240~AU  &		       & Butler et al.\ 2001\\
%                 &&&&	 &&&	       &	  &&  &	     &  		&		       & Chauvin et al.\ 2007\\
%\hline
% SDSS J121209.31$+$013627.7\,b &&&&	 &&&     &&&	    &	     &  		&		       & Farihi et al.\ 2007\\
%\hline
GD\,66\, b (?)   &$M<$7:& 2&.&4  & 2&.&75  & 5&.&7	    & 0      & WD (DAV)         & timing (pulsations)  & Mullally et al.\ 2008, 2009\\
\hline
GD\,356\, b (?)  & $M<$12&&&&    &&         & $>$2&.&7\hfill h  &        & WD, magnetic     & inferred from       & Wickramasinghe et al.\ 2010\\
                 &&&&	 &&&	       &	   && &	     &  		& Zeeman splitting     & \\
\hline
V391\,Peg\,b     && 3&.&2   & 1&.&7         & 3&.&2	    & 0      & EHB (sdBV$_{\!\rm rs}$)	& timing (pulsations)  & Silvotti et al.\ 2007\\
\hline
HW\,Vir\,b   && 19&.&2  & 5&.&3         & 15&.&8	    & 0.46   & EHB (ecl.~sdB+M) & timing (eclipse)     & Lee et al.\ 2009\\
HW\,Vir\,c   && 8&.&5   & 3&.&6         & 9&.&1	    & 0.31   &                  & timing (eclipse)     & \\
\hline
HS\,0705+67003\,b&& 39&.&5  & $<$3&.&6      & 7&.&15	    &	     & EHB (ecl.~sdB+M) & timing (eclipse)     & Qian et al.\ 2009a\\
\hline
HD\,149382\,b (?)&$M=$8--&23&&& 5--6&.&1 \hfill\rsun & 2&.&391\hfill d   &	     & EHB (sdB)	& RVs		       & Geier et al.\ 2009\\
                 &&&&	 &&&	       &	    &&&	     &  		&		       & Jacobs et al.\ 2010\\
\hline
NN\,Ser\,b\tablenote{All numbers given are for the model 2a solution by Beuermann et al. 2010a}&& 6&.&9 & 5&.&4 & 15&.&5 & 0 & pre-CV & timing (eclipse) & Qian et al.\ 2009b\\
NN\,Ser\,c$^{\ast}$&& 2&.&3   & 3&.&4         & 7&.&7	    & 0.2    &                  & timing (eclipse)     & Beuermann et al.\ 2010a\\
                                                                                                  &&&&&&&&&&&&&& Hessman et al.\ 2010\\
\hline
DP\,Leo\,b\tablenote{All numbers given are from Beuermann et al. 2010b}	 && 6&.&05  & 8&.&2 & 28&.&0 &	0.39   & CV (polar) & timing (eclipse) & Qian et al.\ 2010a\\
                 &&&&	 &&&	       &	    &&&	     &  		&		               & Beuermann et al.\ 2010b\\
\hline
QS\,Vir\,b&$M\simeq$&6&.&65\tablenote{If coplanar to the eclipsing binary}&4&.&2&7&.&86&0.37&CV, hibernating& timing (eclipse)& Qian et al.\ 2010b\\
\hline
HIP\,13044\,b    &&1&.&25   & 0&.&116  &    16&.&2\hfill d& 0.25 &  RHB                 &	RVs            & Setiawan et al.\ 2010a\\
                 &&&&	 &&&	       &	    &&&	         &  (extragal.\ orig.?)	&		       & Setiawan et al.\ 2010b\\
\hline
\end{tabular}
\caption{Post-RGB planets}
\label{tab:a}
\end{table}

\clearpage

A circumbinary planet with a minimum mass of about 5~M$_ {\rm Jup}$
could be the reason of the periodic component in the O--C plot of the eclipsing
polar HU~Aquarii (Schwarz et al.\ 2009, not included in Table~1).
O'Toole et al.\ (2010) also report early results from their on-going RV survey.
We refer to their proceeding papers for more details.

For what concerns the white dwarfs, until now there are no clear detections 
of planets around single white dwarfs, with only two candidates:
GD\,66 (Mullally et al.\ 2008, 2009) and GD\,356 (Wickramasinghe et al.\ 2010).
We know at least three planet hosting MS stars having a wide binary
WD companion: HD\,13445=Gl\,86, HD\,147513 and HD\,27442
(Desidera \& Barbieri 2007 and ref.\ therein). %(; see also Mugrauer 2010).
In all these three systems the planet orbit the MS star at an orbital distance
much smaller than the WD distance: 0.11 versus 19~AU for HD\,13445,
1.26 versus at least 100 AU for HD\,147513, and 1.18 versus 240 AU for HD\,27442.
Therefore it is unlikely that the WD had an influence on the evolution
of the planetary system and this is why these systems are not included
in Table~1.
More interesting (although quite exotic) is the case of the circumbinary 
planet orbiting the pulsar+WD system PSR~1620$-$26 (Thorsett et al.\ 1999, 
Sigurdsson et al.\ 2003) in the globular cluster M4.
Following Sigurdsson et al.\ (2003), the most likely scenario is that the
planet was initially in orbit around the MS star progenitor of what is now 
the WD companion.
The system underwent an exchange interaction with a neutron star binary, 
with the MS star replacing the original high-mass WD companion to the neutron
star. 
The end results were a new eccentric orbit for the MS star, and the 
displacement of the planet onto a wide circumbinary orbit.

\section{Ongoing projects and the potential of CoRoT and Kepler data}

Similarly to the astrometric method (and contrary to the RV and transit 
methods), the timing method is sensitive to large orbital separations.
Equation (1) shows that the (maximum) change in the arrival time of the photons (or time 
shift) $\tau$ is indeed proportional to the orbital distance $a$ ($m_p$, 
$M_{\star}$, $c$ and $i$ are, respectively, planet mass, stellar mass, light 
speed and inclination of the orbit to the line of sight):

\begin{equation}
\tau~=~\frac{a}{c}~~\frac{{m_p\,\,{\rm sin}\,i}}{M_{\star}}
\end{equation}

\vspace*{4mm}

\noindent
For\, $a$\,=\,1\,AU, $m_p$\,=\,1\,M$_{\rm Jup}$, 
$M_{\star}$\,=\,1\,\msun\,(and $i$=90$^\circ$),
the time shift is about 0.5 seconds; while it is 1\,s for a stellar
mass of 0.5\,\msun, as in the case of a typical sdB star or a slightly 
lower-than-average-mass white dwarf.

As shown by the recent discoveries in the previous section, it is possible to
detect planets with the timing method using ground-based data.
What is needed is a long-term photometric monitoring of the targets for 
many years with frequent runs, ideally each month or two.
A few ground-based programs to search for post-RGB planets exist, e.g.\
the EXOTIME program (Schuh et al.\ 2009, Schuh 2010, Lutz et al.\ 
2010\footnote{see also {\url{http://www.na.astro.it/~silvotti/exotime/}}}) 
which is monitoring five sdB pulsators, a similar program at the McDonald 
Observatory on 15 ZZ~Ceti pulsators \cite{Mullally+2008}, and
the long-term monitoring of one DBV pulsator by Dalessio et al.\ (2010).
A secondary goal of these programs is the measurement of the secular variation
of the pulsation periods, which is related to the change of the stellar 
structure because of its evolution (e.g.\ Kepler et al.\ 2005, 
Costa \& Kepler 2008, Silvotti et al.\ 2007).
Similar ground-based programs exist also for the eclipsing binaries.

However it is clear that, potentially, the maximum results can be achieved
using continuous monitoring from space craft.
The exceptional photometric precision and duty cycle of the CoRoT and Kepler 
satellite data represent a unique opportunity to use the timing method to detect low mass
substellar companions to pulsating stars and eclipsing binaries.
In particular, Kepler has a longer coverage, up to 3.5 years, as the satellite
will observe the same $\sim$105 square degree field for the whole mission 
duration; while the CoRoT ``long runs'' have a maximum duration of about 6 
months.
For more details concerning the Kepler and CoRoT missions we refer the reader 
to Borucki et al.\ (2010), Gilliland et al.\ (2010) and Auvergne et al.\ (2009).
The main limitation of Kepler is the relatively long sampling time of 58~s 
(short-cadence data).
Although the photometer samples the field every 6.54 s, telemetry restrictions 
do not permit to save and transmit to the ground the original data.
This is a severe limitation only for the very rapid pulsators like the 
short-period sdB stars of the sdBV$\!_{\rm r}$ class and the white dwarfs, which can
have pulsation periods as short as $\approx$1--2~minutes.
However, in the survey phase (1st year of observation) that has just ended, 
no pulsating white dwarfs were found and only one short-period sdB pulsator was
observed (Kawaler et al.\ 2010) plus an hybrid pulsator with a very rich 
amplitude spectrum in a eclipsing binary system with a M dwarf
(\O stensen et al.\ 2010a, see also end of this section on {\it5. Eclipse timing}).

For all the other 13 long-period sdB pulsators (sdBV$\!_{\rm s}$) discovered by Kepler 
(\O stensen et al.\ 2010b, 2010c),
with typical periods between 30 min to few hours, the 58~s sampling is largely
sufficient, even though longer pulsation periods mean higher minimum detectable
mass of a substellar companion.
This is shown by the following equation (2), in which $\sigma_\tau$ is the 
timing error, $N$ is the number of data points,
$P$ the pulsation period, $\sigma_I$ the photometric (relative intensity)
error, $A$ the pulsation amplitude and $\sigma_A$ the amplitude
error, which decreases with the square root of the length of the run and
can be estimated directly by measuring the mean noise of the Fourier 
amplitude spectrum.
Equation (2) is valid under the assumption of uncorrelated errors and
no cross terms between the different pulsation frequencies, and is obtained
multiplying by $P$ equation (A15) from Breger et al.\ 1999 
(see also Montgomery \& O'Donoghue 1999 and Silvotti et al.\ 2006 for more 
details).

\begin{equation}
\sigma_\tau=\frac{P}{2\pi}~\sqrt{\frac{2}{N}}~\frac{\sigma_I}{A}
=\frac{P}{2\pi}~\frac{\sigma_A}{A}
\end{equation}

\vspace*{4mm}

\noindent
A similar equation can be obtained also for the eclipse timing 
(Doyle \& Deeg 2002), assuming a simple model of triangular eclipse
(which implies that the error in eclipse minimum timing $\sigma_\tau$ is a 
lower limit to the real error).
In the following equation (3), $T_{\rm ecl}$ is the duration of the eclipse
from first to last contact, $\Delta I$ is the relative intensity of
the eclipse (or (1-$\Delta I$) is the eclipse's relative depth, 
with $I$=1 out of eclipse) and $N_{\rm ecl}$ is the number of data points
taken during $T_{\rm ecl}$; $\sigma_I$ was defined above.

\begin{equation}
\sigma_\tau=\frac{T_{\rm ecl}}{2\sqrt{N_{\rm ecl}}}~\frac{\sigma_I}{(1-\Delta I)}
\end{equation}

\vspace*{4mm}

As we do not expect to find close-in planets with short orbital periods
around post-RGB stars\footnote{And in any case the timing method is not
very sensitive to small orbital distances, see equation (1).},
we will need many months/years of Kepler data to be able
to set limits on the presence of planets around these stars.
At the moment, with the available Kepler data, we can only start making some 
tests to see what the typical phase errors are that we can obtain and therefore
what we can expect in terms of minimum detectable masses and orbital periods.
The first goal of these preliminary tests is to verify the phase stability
(or ``phase coherence'') of different classes of pulsating stars and eclipsing 
binaries.

Regarding the pulsators, the most promising targets are those with stable 
pulsations (in terms of period, amplitude and phase), short periods 
(to reduce the minimum detectable mass), 
several independent periods and relatively large amplitudes.
To have several independent periods is important as it permits to check whether
the same periodic pattern produced by a planet is present in all the periods 
(see Dalessio et al.\ 2010 for an example where this is not the case\,!),
helping to exclude false detections due to other periodic effects (e.g.\ beating
of close frequencies).
SdB pulsators fit well these constraints and, due to the relatively short 
periods, are the best candidates.
Other potential targets are the Delta Scuti stars, $\beta$ Cephei and RR Lyrae.
In the next pages we show the results of a few preliminary tests 
on these pulsating stars using Kepler and CoRoT data, including also
an eclipsing binary with a pulsating sdB component.
The Kepler data were retrieved from the KASC data base (KASC is the Kepler 
AsteroSeismic Consortium, Gilliland et al.\ 2010).
The CoRoT public data were downloaded from the CoRoT archive at LAEFF
(Laboratory for Space Astrophysics and Theoretical Physics) in Spain.

\vspace{4mm}
\noindent
{\bf 1. Subdwarf B stars}
\vspace{4mm}

\noindent
Before the Kepler observations, the oscillating sdB stars could be divided into
two groups: the short-period (p-mode) pulsators (or sdBV$\!_{\rm r}$=rapid pulsators)
of the V361 Hya class from the prototype (Kilkenny et al.\ 1997),
with effective temperatures between about 27,000 and 35,000 K, 
and the longer period (g-mode) pulsators (or sdBV$\!_{\rm s}$=slow pulsators)
of the V1093 Her class (Green et al.\ 2003), 
with effective temperatures between $\sim$21,000 and $\sim$28,000 K.
Four hybrid pulsators (sdBV$\!_{\rm rs}$) were known, showing both p- and g-modes at the same 
time (Schuh et al.\ 2006, see also Baran \& Fox-Machado 2010 
for the complete list of hybrid pulsators before Kepler).
This classification (compare also Kilkenny et al.\ 2010)
may need some adjustment because the Kepler observations
have shown that the temperature range of the two instability strips and the 
fraction of hybrid pulsators could be larger than previously believed.

In order to have a first feeling of what can be obtained with the timing 
technique on Kepler sdB data, we consider the star KIC\,010139564, the only 
rapid sdB pulsator observed by Kepler \cite{Kawaler+2010}, and one of the best 
candidates to search for planets.
Figure~1 shows the O--C plot of six pulsation frequencies: these frequencies are
considered also in Figs.\,4 and 5 of Kawaler et al.\ (2010).

\begin{figure}[t]%[hpt]
\includegraphics[width=\textwidth]{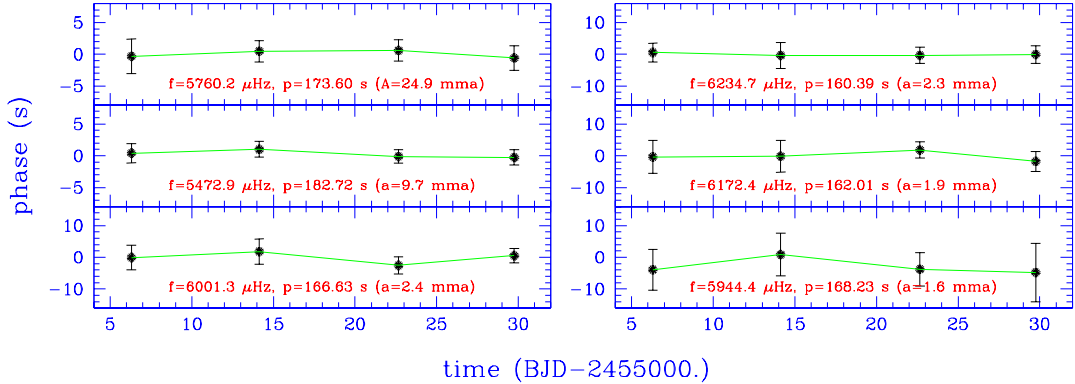}
\caption{O--C plots of 6 pulsation periods of the short-period sdB pulsator 
KIC\,10139564 (see Kawaler et al.\ 2010 for more details).
Each point represents about 7.5 days of short cadence data.
The phase stability of all these periods suggests that substellar companions
can be detected.
The small error bars of the two periods with higher amplitude 
(173.60 and 182.72 s) allow to detect a planet of about 7 M$_{\rm Jup}$ at 1 
AU from the star.
Using longer subsets, this limit can be reduced to about 2 M$_{\rm Jup}$ and
even further (see text for more details).}
\label{f1}
\end{figure}

First of all, we see from Fig.\,1 that the frequencies have in general stable 
phases.
This is true not only for these six frequencies but also for almost all the 
other frequencies of this star, although they are not represented here.
By comparing Fig.\,1 with Fig.\,4 of Kawaler et al.\ (2010), we see that, 
for most of these frequencies, the error bars are smaller by a factor of 
(7.5/2)$^{0.5}\simeq$1.9, which is exactly what we expect considering that 
each point in Fig.\,1 represents $\sim$7.5 days of data while in Kawaler's 
plot each point corresponds to 2 days.
However, at least one frequency at 5760.2~$\mu$Hz behaves differently: the 
error bar of Fig.\,1 is larger than the one obtained by Kawaler et al.\ (2010).
The reason is in their Fig.\,5, where we see that both phase and amplitude are
changing periodically with a phase difference of about 90 degrees, suggesting
that beating between closely spaced stable oscillation frequencies is 
occuring.
From the smaller error bars of Fig.\,1 we can estimate the minimum mass of
a detectable planet: we obtain about 7 M$_{\rm Jup}$ for an orbital distance of
1 AU.
However, potentially, the minimum mass can be reduced to about 2 
M$_{\rm Jup}$ using longer subsets of about 3 months each. This limit could be 
further extended to about 0.7 M$_{\rm Jup}$ extrapolating from the small error 
bars in Fig.\,4 of Kawaler et al.\ (2010), provided that the new Kepler data 
that will be collected in the next months/years will actually allow to 
separate
the close frequencies and resolve the amplitude spectrum of this star.

For the long-period sdB pulsators, the first observations have shown that in 
general the phases are stable allowing to use the timing method also for these
stars.
For them, the minimum detectable mass of a substellar companion should be 
$\approx$10--50 times higher than the numbers reported above considering that 
the periods are at least one order of magnitude longer.
However, some of these stars show also a few peaks at high frequency, that are 
probably due to a few excited p-modes (Reed et al.\ 2010, Charpinet et al.\ 2011). 
If these short-period p-modes will be confirmed by the future Kepler 
observations, they could bring down again the minimum detectable masses of the
potential companions.

\vspace{4mm}
\noindent
{\bf 2. Delta Scuti stars}
\vspace{4mm}

\noindent
Another interesting class of pulsators are the Delta Scuti stars.
Although the periods are relatively long and comparable with those of
the long-period sdB pulsators, their amplitudes can be larger,
allowing to reduce the timing errors to a few seconds (see equation 2).
An example of a Delta Scuti star with a relatively simple amplitude spectrum
(not too many frequencies) and short periods is given in Fig.\,2, 
where we see that the phases are stable, at least during the
1-month Kepler run in short cadence.

\begin{figure}[t]%[ht]
\includegraphics[width=\textwidth]{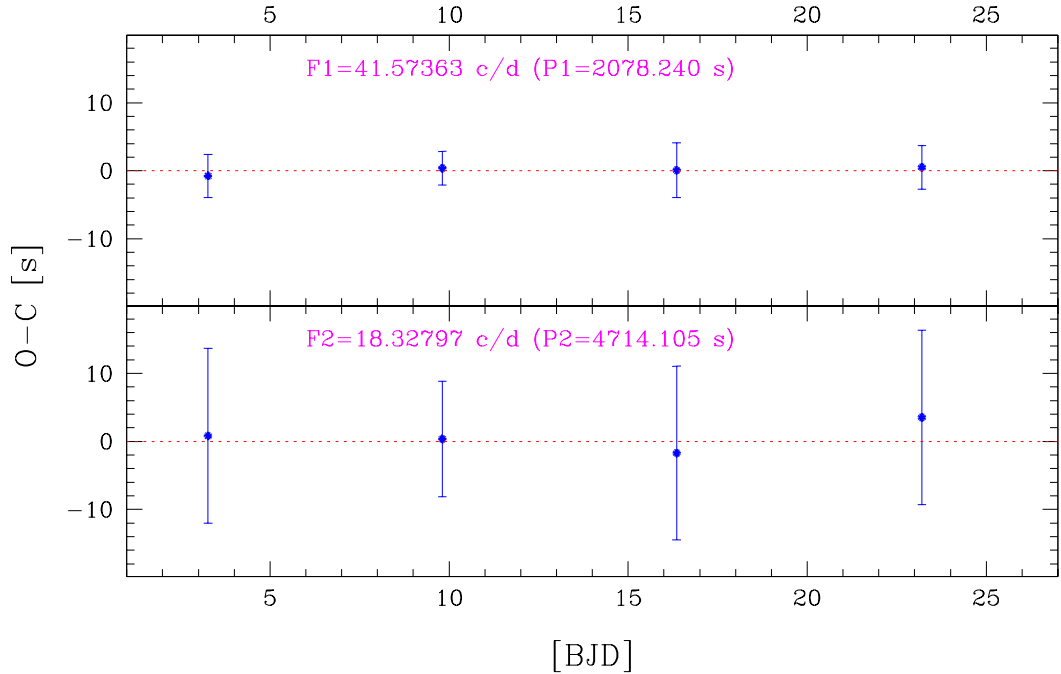}
\caption{An example of O--C plots of a Delta Scuti star observed by Kepler.
Each point represents about 7 days of short cadence data.
The phase errors of the two main pulsation frequencies F1 and F2,
of the order of $\pm$4 and $\pm$10~s respectively, suggest that
Delta Scuti stars can be stable enough to use the timing technique.
The fractional amplitudes of the two frequencies are 0.25\% and 0.17\% 
respectively.}
\label{f2}
\end{figure}

\vspace{4mm}
\noindent
{\bf 3. $\beta$ Cephei stars}
\vspace{4mm}

\noindent
The timing technique can also be used for $\beta$ Cephei stars.
Here we show an example on the $\beta$ Cephei star HD~180642 (V1449 Aquilae),
illustrating that these stars also may have stable phases.

HD~180642 was observed by the CoRoT satellite during 156 days of the
long run LRc01, with a sampling cadence of 32\,s.
The CoRoT data were analyzed in detail by Degroote et al.\ (2009), who found
a complex amplitude spectrum with several combination frequencies.
We applied a least-square sinusoidal fit to the CoRoT data using 15
frequencies obtained from pre-whitening, including a few combinations.
In Figure~3 (upper panel) a preliminary O--C plot shows an apparent phase 
variation of the order of 20\,s in 2 months, whose origin and nature is not clear.
If this effect were due to a secular variation of the pulsation period, we 
would obtain \pdot\ $\approx$ $4\times$10$^{-9}$ s/s.
This number is significantly larger than the value found by Degroote et al.\ 
(2009) from direct measurement: about 1.3$\times$10$^{-9}$ s/s. 
This difference deserves further study.

\begin{figure}[t]%[!h]
\includegraphics[width=\textwidth]{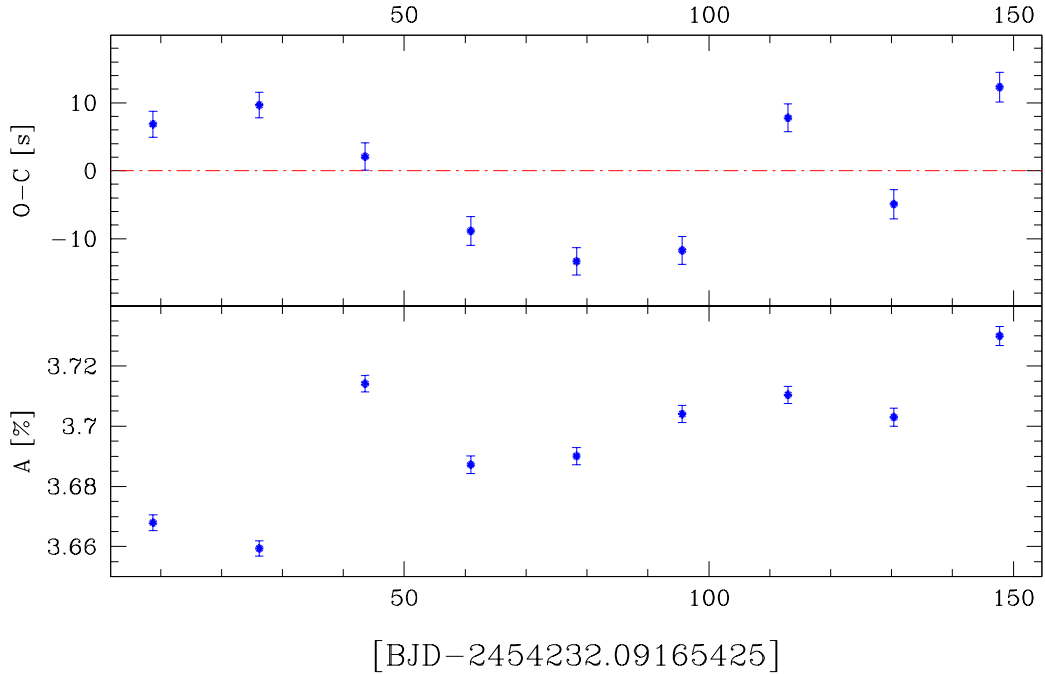}
\caption{The upper and lower panels represent the phase and 
(fractional) amplitude variations, respectively, of the main pulsation frequency 
F=5.486889 c/d (P=4.374063 h) of the $\beta$ Cephei star HD~180642 (V1449 
Aquilae).
Each point represents about 17.4 days of the CoRoT long run LRc01 (156 days).
The two panels suggest that phase and amplitude variations are not correlated
and therefore these variations can not be attributed to beating 
of close frequencies.
See Degroote et al.\ (2009) for more details on the pulsational behavior of 
this star.}
\label{f3}
\end{figure}

\vspace{4mm}
\noindent
{\bf 4. RR Lyrae stars}
\vspace{4mm}

\noindent
From the KASC data base we selected two RR Lyrae stars with the longest 
observational coverage (about 6 months).
The O--C diagram of these RR Lyrae stars, shown in Fig.\,4, indicates that
the phase errors are generally large and, even more important, they are 
highly inhomogeneous.
The reason of this behaviour is not clear.
Although before the Kepler observation, both stars were considered as 
non-Blazhko stars, the non-uniform error distribution might be due to a 
low-level Blazhko modulation of phases and/or frequencies.
Noise due to convection might also play a role.
Although a deeper testing on a larger number of RR Lyrae stars is needed 
before drawing conclusions, this first example suggests that RR Lyrae might 
not be stable enough for the timing technique.

\begin{figure}[t]%[!h]
\includegraphics[width=\textwidth]{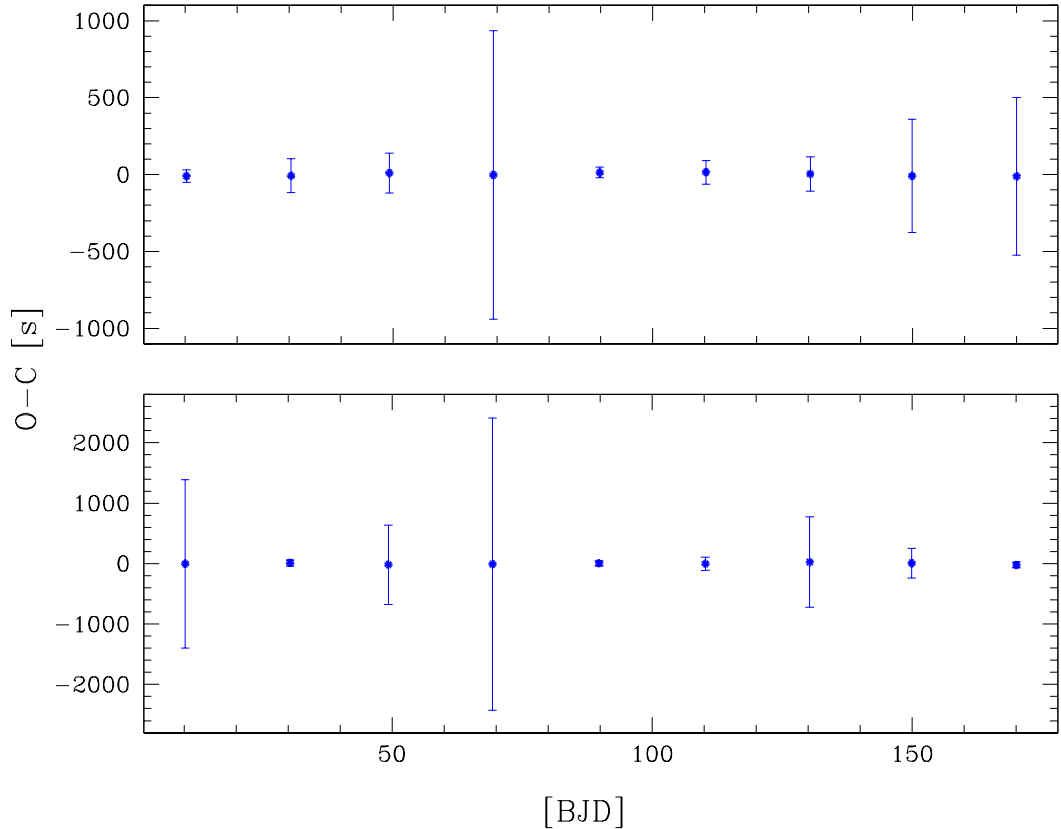}
\caption{O--C diagram of two RR~Lyrae stars observed by Kepler in long 
cadence (a measurement each 30 min). 
Each point represents about 20 days of data.
The vertical scale is different in each panel.
The reason of the large and inhomogeneous phase errors is not clear
(see text for more comments).}
\label{f4}
\end{figure}

\vspace{4mm}
\noindent
{\bf 5. Eclipse timing}
\vspace{4mm}

\noindent
In this section we show an example of eclipse timing. 
We note that this method is related to the search for transit timing variations.
The target, 2M1938+4603 (or KIC\,09472174), is of particular interest because 
it is a sdB+dM eclipsing binary system and the sdB component shows a very rich
pulsational spectrum with periods ranging from $\sim$4 min to 5.5 hours.
The orbital period of the system is 0.12576530 d.
For more details on 2M1938+4603 we refer the reader to the recent article 
by \O stensen et al.\ (2010a).
The eclipsing binary systems with a pulsating component, like this one, have 
the advantage of having two independent clocks given by the eclipse timing 
and by the pulsations (provided that there is no coupling between them).
For this reason they are an ideal test 
case to verify the potential of the timing method in finding low-mass 
substellar objects.

\begin{figure}[!t]
\includegraphics[width=\textwidth]{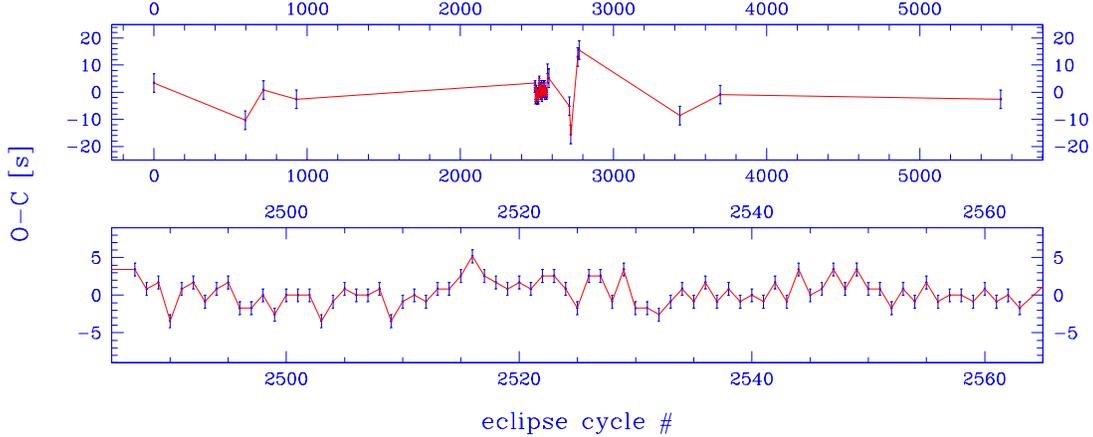}
\caption{O--C diagram of the eclipsing sdB+dM binary 2M1938+4603 
(KIC\,09472174), based on eclipse timing. The times of the primary minima are 
taken from \O stensen et al.\ (2010a).
With a binary period of 0.12576530 d, the upper panel corresponds to 695 days
of ground based eclipse timings, with the Kepler 9.7-day commissioning data 
around cycle 2500.
The latter are shown in more detail in the lower panel. 
The noise in the lower panel (and the detectable planetary mass) can be 
reduced by fitting several eclipses at once.
To have an idea of the realistic timing errors, if we consider all the 77
Kepler eclipses in the lower panel, we obtain an O--C standard deviation of 
1.8~s.
With a 2 years (or longer) observation, and using 2.5 months of data for each 
point in the O--C plot, the 4 $\sigma$ detection limit would be reduced to 
about 4 M$_{\rm Jup}$ considering an orbital distance of 1 AU from the binary.
Potentially, the presence of a planet could be independently 
confirmed also from the pulsations of the sdB star.}
\label{f5}
\end{figure}

\section{Conclusions and future perspectives}

In the last three years, the first discoveries of planets orbiting 
evolved stars after the red giant branch have proven that the timing method
is an efficient way to detect substellar objects around compact stars,
using either the stellar pulsation or the eclipse timing as a clock.
This method, which is sensitive to Jupiter-mass planets in large orbits 
of $\approx$1 AU or more, requires long-time photometric monitoring with
frequent, or, better still, continuous, runs.
In order to validate the method, it would be important to confirm these
discoveries with other independent techniques.
This in general is not easy because the targets are generally faint and 
many have spectral lines broadened by high gravity.

The high quality data from the CoRoT and Kepler satellites represent
a unique opportunity due to the exceptional duty cycle.
Kepler has the additional advantage of longer runs, up to 3.5 years.
Preliminary tests show that sdB stars are the best candidates, in particular 
the short-period pulsators.
Other good candidates are the Delta Scuti stars, at least those with a 
relatively simple amplitude spectrum (not too many frequencies).
An example on a $\beta$ Cephei star shows that also these stars may have
sufficiently stable phases, although the higher stellar mass implies a higher 
minimum mass of the potential substellar companions.
RR Lyrae stars are more problematic and our preliminary tests suggest that phases 
and/or frequencies may have significant variations on time scales of weeks.
An example of an eclipsing binary is given and shows that also for these 
targets Jovian planets can be detected.
Moreover, if the binary system has a pulsating component like 2M1938+4603,
the pulsations can be used, potentially, as an independent clock to confirm
the detections.

\begin{theacknowledgments}
R.S. acknowledges financial support from the organizers, who
waived the conference fees.
\end{theacknowledgments}



\bibliographystyle{aipproc}   % if natbib is available


% \bibliography{sample}


%  \typeout{******************************************}
%  \typeout{** Please run "bibtex \jobname" to optain}
%  \typeout{** the bibliography and then re-run LaTeX}
%  \typeout{** twice to fix the references!}
%  \typeout{******************************************}

%\end{document}

%%%%%%%%%%%%%%%%%%%%%%%%%%%%%%%%%%%%%%%%%%%
%% The following lines show an example how to produce a bibliography
%% without the help of the BibTeX program. This could be used instead
%% of the above.
%%%%%%%%%%%%%%%%%%%%%%%%%%%%%%%%%%%%%%%%%%%

\end{document}